\documentclass[twocolumn,showpacs,preprintnumbers,amsmath,amssymb,floatfix,showkeys]{revtex4}

\usepackage{graphicx}
\usepackage{dcolumn}
\usepackage{bm}

\def\Includegraphics#1 {Here, include graphics #1}            

\def \D {\hbox{d}}
\def \cotg  {\mathop{\rm cotg}\nolimits}
\def \sech{\mathop{\rm sech}\nolimits}
\def \barA {\overline{A}}

\def \ee  {     \mathcal{E}}  
\def \ec  {\bar{\mathcal{E}}} 
\def \intIm           {I_{\rm m}}
\def \intIc {\overline{I_{\rm m}}}
\def \intId           {I_{\rm d}}

\def\netgain {net gain}

\def\gnl{\gamma_{\rm NL}}

\def \mod#1{\vert #1 \vert}

\def\vg{v_{\rm g}}
\def\vc{\overline{v}_{\rm g}}

\def\today{Accepted 11~November~2009, Physical Review E}

\begin{document}

\preprint{S2009/???}

\title{
Ginzburg-Landau equation for dynamical four-wave mixing in gain nonlinear 
media with relaxation
}

\author{Svitlana Bugaychuk,$^{1,}$} 
\email{bugaich@iop.kiev.ua}
\author{Robert Conte,$^{2,}$}
\email{Robert.Conte@cea.fr}

\affiliation{
$^1$Institute of Physics,
National Academy of Sciences,
46 Prospect Nauki, Kiev 03028, 
Ukraine,
}

\affiliation{
$^2$LRC MESO,\'Ecole normale sup\'erieure de Cachan (CMLA) et CEA--DAM
\\ 61, avenue du Pr\'esident Wilson, F--94235 Cachan Cedex, France.
\\
Service de physique de l'\'etat condens\'e (CNRS URA 2464),
CEA-Saclay, F-91191 Gif-sur-Yvette Cedex, France
}

\date{\today}

\begin{abstract}
We consider the dynamical degenerate four-wave mixing (FWM) model 
in a cubic nonlinear medium including both the time relaxation 
of the induced nonlinearity and the nonlocal coupling. 
The initial ten-dimensional FWM system can be rewritten 
as a three-variable intrinsic system (namely the intensity pattern, 
the amplitude of the nonlinearity and the total net gain)
which is very close to the pumped Maxwell-Bloch system. 
In the case of a purely nonlocal response the initial system reduces 
to a real damped sine-Gordon (SG) equation. 
We obtain a new solution of this equation 
in the form of a $\sech$ function with a time-dependent coefficient.
By applying the reductive perturbation method to this damped SG equation,
we obtain exactly the cubic complex Ginzburg Landau equation (CGL3), 
but with a time dependence in the loss/gain coefficient.
The CGL3 describes the properties of the spatially 
localized interference pattern formed by the FWM.

\end{abstract}


\maketitle

\section{Introduction}
\label{sec:Introduction}

The effect of interaction of light and matter in nonlinear optics 
is very often characterized by a coupling coefficient 
which reveals a response of the matter.
If in addition a mutual mixing of several waves is taken into consideration, 
one deals with nonuniform spatial (or spatiotemporal) fields. 
People usually consider the reaction of matter is local on the action of the field.
But this is not always the case.
In inertial or nonlocal systems the response can be retarded in time or shifted in space.
As a result, the beam-coupling coefficient takes 
a complex value 
and some phase addition appears between the mixed waves.  
This can lead to the control of parameters of one beam 
by guiding the properties of another beam, 
as well as to the formation of stable localized structures (i.e.~intensity patterns).
In this paper we show rigorously that a nonlinear system 
describing the degenerate wave mixing in a medium 
which possesses both a nonlocal response and relaxation 
is reduced to one nonlinear complex Ginzburg-Landau equation (CGLE). 
We develop the technique to obtain the cubic CGLE by using 
the reductive perturbation method for the nonlinear dynamical 
wave coupling system.

Nonlinear dynamical systems have been studied intensively during the last 
decennia after localized structures (e.g.~solitons) 
were found in such systems. 
The complex Ginzburg-Landau equation (CGLE) became a widely used 
physical-mathematical model appearing in many branches of physics, chemistry 
and biology, in order to describe various localized structures 
\cite{AK2002,vS2003,AABook2005,AABook2008}. 
Moreover the CGLE is considered as the simplest model containing dissipative 
soliton solutions, which exist in nonequilibrium systems where gain and loss 
are balanced \cite{AABook2005,AABook2008,RosanovBook}. 
In optics, the dissipative solitons described by the Ginzburg-Landau equation 
appear for pulsed operation of passively mode-locked lasers 
as well as for all-optical long-haul soliton transmission lines 
\cite{AABook2005,AABook2008,AkSotoPRE2001,AKHTPRL2008}. 

The dissipative models which take into account wave interactions have been 
studied in \cite{Montes,CDT},
first of all as the envelope of dissipative solitons emitted by an optical 
parametric oscillator.
In \cite{Montes} the author presents theoretical and experimental studies of 
stumilated Brillouin back-scattering of a continuous pump wave
resulting in backward-traveling solitary pulses in long fiber-ring cavities. 
Nonlinear optical cavities with three-wave interaction in a nonlinear crystal,
when the waves have different frequencies, 
were considered in Ref.~\cite{CDT}.
It was shown that the spatial dissipative solitons can form spontaneously 
in that case.
For the first time we consider the cubic CGLE which appears in the problem 
of dynamical interaction of four waves with the same frequencies 
in extended nonlocal media.
We show that the CGLE is obtained because of a photo-induced nonlocal 
nonlinear response 
which includes a time relaxation term in the considered (dissipative) model.

The next feature that we utilize in the model, the nonlocality,
reveals itself as an ubiquitous 
property in many branches of physics, 
e.g. optics, plasmas, Bose-Einstein condensates \cite{BECNL,KonotopPRE2000}.
Usually the nonlocal response appears 
when the nonlinearity is associated with some sort of transport process 
such as heat conduction in media with thermal response \cite{RCMSPRL2005}, 
diffusion of molecules or atoms accompanying nonlinear light propagation 
in atomic vapors \cite{SuterBlasberg,SkupinPRL2007}, 
and charge transport in photorefractive crystals \cite{ZozAnd1995,TornerPRL2005}.
Specific properties of spatial solitons were investigated 
in nematic liquid crystals, 
where nonlocal response exists due to reorientation of anisotropic molecules 
by a propagating beam \cite{Krolikowski2001,CPAPRL2003,KivsharPRA2007,Guo2008}.
The nonlocal nonlinearities with formation of 
dissipative optical solitons for a wide-aperture laser with saturable absorption 
were studied recently in \cite{LedererAABook2005,RosanovAAB2008}.

One usually investigates stationary changes of the induced nonlinearity. 
Our dissipative model includes both gain and relaxation of the nonlinearity in a 
nonlocal medium. 
Since we consider the process of wave coupling, 
the photoinduced nonlocal nonlinearity leads to an effect of energy transfer 
between waves during their propagation. 
In this way the nonlocality plays the role of an amplified medium to increase 
the intensities of one beam at the cost of decreasing the energy of another beam. 
The energy transfer effect is observed in the dynamical holography 
when the interacting beams record a dynamical grating, 
which is shifted from the interference pattern, 
and the same beams diffract from this grating \cite{TAP61,Trends87}.
As a result of this energy transfer both the interference pattern 
and the spatial distribution of the amplitude of the nonlinearity 
get a stable localized pattern along 
the $z$-longitude direction of the medium \cite{HongSaxema,JBH_JOSAB,BKMPR}.
We show that the CGLE governs the spatio-temporal dynamics for both values.

Dissipative solitons described by CGLE demonstrate a rich variety 
of unusual properties \cite{AABook2005,Tsoy2006}, 
such as stable periodic pulsations, bounded solitary waves, 
periodic ``explosions'', and collapse.
All these unique features may find applications in nonlinear wave coupling,
in particular in the dynamic holography in media with nonlocal response. 
Among possible applications in photonics let us mention: 
(i) holographic interferometers including phase-shifted interferometers; 
(ii) traps of light (trapping states) in a resonator; 
(iii) manipulation of pulses having different intensities and durations 
in order to obtain optical logic elements, all-optical switching, pulse retardation etc.,
as well as the interaction of pulses not only in bulk materials 
but with thin nonlinear films, nanomaterials and metamaterials; 
and many others.
During the process the medium should possess a nonlocal nonlinearity, 
e.g.~some kind of transport mechanism; 
or 
the medium can have a local nonlinearity 
but a regime of moving dynamical gratings should be realized.

The paper is organized as follows.
In section \ref{sec:II} we introduce the four-wave mixing model
and recall the existing results.
In section \ref{section_dampedSG},
we revisit the derivation of the damped sine-Gordon equation
and derive a new solution to the FWM.
Finally, 
in section \ref{section_multiscale},
we apply the method of multiple scale expansion 
and find as a result the cubic CGLE.
This procedure proves that the FWM as well as the dynamical 
self-diffraction of waves can be considered as a dissipative nonlinear system
containing stable soliton solutions.

\section{The intristic system of the dissipative FWM model}
\label{sec:II}

The one-dimensional degenerate FWM initial system consists of five partial 
differential equations, namely, 
four coupled wave equations for slow variable amplitudes 
which connect waves 1 and 2 propagating 
in a forward direction and waves 3 and 4 propagating in 
a backward direction,
\begin{eqnarray}
& &
\partial_z     A_1=-i \ee     A_2,\
\partial_z \barA_2= i \ee \barA_1,\
\nonumber\\ & &
\partial_z \barA_3=-i \ee \barA_4,\
\partial_z     A_4= i \ee     A_3,\
\label{eqFWM}
\end{eqnarray}
and the dynamical equation for the medium, which in the simplest case includes only 
a gain being proportional to the intensity pattern and an exponential relaxation, 
in the form 
\begin{eqnarray}
& &
     \partial_t \ee = \gamma \intIm - \frac{\ee}{\tau}, 
\label{eqgrating_t}
\end{eqnarray}
We assume here that the interference pattern is formed by two pairs of 
co-propagating waves
\begin{eqnarray}
& &
\intIm =A_1 \barA_2 + \barA_3 A_4.
\label{eqdefIm}
\end{eqnarray}

In Eqs.~(\ref{eqFWM})--(\ref{eqdefIm}),
$A_j(t,z)$ is the slow variable amplitude
of the $j$-th plane wave 
$E_j(t,z)=A_j(t,z) e^{i(wt-\vec{k}_j \vec{r})}$, 
$\ee(t,z)$ is the amplitude of the photoinduced nonlinear susceptibility. 
It must be emphasized that the response constant
$\gamma=\gamma_{\rm L} + i \gnl =\mod{\gamma} e^{i g}$ 
is complex.
The complex value of the coupling coefficient $\ee$ 
is an essential feature for the existence of soliton-like solutions. 
The interacting waves are connected by the impulse conservation law:
\begin{eqnarray}
& &
\vec{k_1}-\vec{k_2}=\vec{k_4}-\vec{k_3}.
\end{eqnarray}
We assume the following normalization: 
all wave amplitudes are normalized by the square root of the total light intensity 
$I_0=\left|A_1\right|^2+\left|A_2\right|^2+\left|A_3\right|^2+\left|A_4\right|^2=
\hbox{ const}$,
$\ee$ is the dimensionless coefficient of the nonlinearity, 
$z$ is the dimensionless longitudinal coordinate $z=(k_0^2/(2 k_z')) z'$, 
where $k_0$ is the amplitude of the wave-vector in the free space,
$z'$ is the spatial coordinate.  
We keep the dimension of the time-coordinate $t$ in order to display
the dependence of the 
 dispersion relation on the time relaxation constant $\tau$.
In this way, in order to make Eq.~(\ref{eqgrating_t}) dimensionless,
the gain coefficient is normalized by the time relaxation constant $\tau$ and 
has the dimension $\left[\gamma \right] = T^{-1}$. 

The system (\ref{eqFWM})--(\ref{eqdefIm}) has been considered
for the dynamic holography 
in the case of a purely 
nonlocal response $\gamma = i \gnl$. 
Then $\ee$ is interpreted as the amplitude of the dynamical grating.
As previously found 
\cite{Zozulya,BKK,JBH_JOSAB,BKMPR},
the initial system is then reducible to a damped sine-Gordon equation (SG), 
which has a stationary solution in the form of a $\sech$ function 
$\mod{\ee} = \gamma C / \cosh \left[2 \gamma C z - p \right]$, 
with $C, p$ arbitrary constants.
Numerical solutions in the form of periodic oscillations were investigated in 
\cite{BKK}.
The first experimental observation of localization of the dynamical grating amplitude 
along the longitudinal coordinate in bulk ferroelectric crystals was made
in \cite{BKMPR}.
For the general case of a complex $\gamma$,
the general stationary solution was later found in \cite{CB2009a},
together with, 
in the dynamical case,
the general solution (expressed with elliptic functions)
of the reduction $(z,t) \to \sqrt{z} e^{-t/\tau}$
for a purely nonlocal response.

The ten-dimensional system
(\ref{eqFWM})--(\ref{eqdefIm})
is invariant 
under any time-dependent rotation
in the space 
$\lbrace A_1,\overline{A}_2,A_4,\overline{A}_3\rbrace$
which preserves the interference pattern (\ref{eqdefIm}).
In a previous work \cite{CB2009a}, 
we could remove
this five-parameter unessential freedom
and obtain the following intrinsic system,
\begin{eqnarray}
& &
\partial_z \intIm = - i \ee \intId,\
\partial_z \intId=- 2 i \ec \intIm + 2 i \ee \intIc,\
\nonumber\\ & &
\partial_t \ee = \gamma \intIm - \frac{\ee}{\tau},
\label{eqFWM555} 
\end{eqnarray}
admitting the first integral
\begin{eqnarray}
& &
4 \mod{\intIm}^2 + \intId^2 = K(t),\ K \hbox{ arbitrary}.
\label{eqFI0}
\end{eqnarray}
The real field $\intId$ 
\begin{eqnarray}
& &
\intId=-\mod{A_1}^2+\mod{A_2}^2-\mod{A_3}^2+\mod{A_4}^2,
\label{eqdefId}
\end{eqnarray}
is the relative \textit{\netgain}.
Therefore the four-wave mixing is characterized by
three intrinsic variables:
the intensity pattern $\intIm$,
the grating amplitude $\ee$
and the relative \netgain\ $\intId$.

This intrinsic system (\ref{eqFWM555}) 
is very similar to 
the pumped Maxwell-Bloch system,
an integrable system of nonlinear optics
defined as
\cite{BZM1987} 
\begin{eqnarray}
& &
\partial_X \rho= N e,\
\partial_X \overline{\rho}= N \overline{e},\
\nonumber\\ & &
\partial_X N=-(\rho \overline{e} + \overline{\rho} e)/2 + 4 s,\
\nonumber\\ & &
\partial_T e=\rho,\
\partial_T\overline{e}=\overline{\rho},\
\label{eqMaxwellBlochComplex}
\end{eqnarray}
with $s$ a real constant (the system is ``pumped'' when $s$ is nonzero).

When the four-wave mixing model is undamped
($\tau=+\infty$)
and has a purely nonlocal response ($\Re(\gamma)=0$),
while the Maxwell-Bloch system is unpumped ($s=0$),
these two systems can be identified,
\begin{eqnarray}
& & {\hskip -12.0 truemm}
\frac{1}{\tau}=0,\
\Re(\gamma)=0,\
s=0:\ 
\nonumber\\ & & {\hskip -12.0 truemm}
 \frac{z}{X}
=\frac{t}{T}
=\frac{2 \mod{\gamma} \intIm}{\rho}
=\frac{2 \mod{\gamma} \intIc}{\overline{\rho}}
=\frac{  \mod{\gamma} \intId}{N}
=\frac{- 2 i \ee}{e}
=\frac{  2 i \ec}{\overline{e}},
\label{eqIntrinsic_identical_MaxwellBloch}
\end{eqnarray}
and in this case the undamped, purely nonlocal response four-wave mixing model 
admits all the solutions of the unpumped complex Maxwell-Bloch system.

\section{Derivation of the damped sine-Gordon equation}
\label{section_dampedSG}

As shown in \cite{BKK,Zozulya,JBH_JOSAB,BKMPR},
under some specific assumptions,
the system made of the four complex equations
(\ref{eqFWM}) can be integrated explicitly.
Because we need it later, let us first establish this derivation
in full generality.

If one represents the complex amplitudes as
\begin{eqnarray}
& & {\hskip -12.0 truemm}
A_j=M_j e^{i \varphi_j},\
\ee=M_e e^{i \varphi_e},\
(M_j,M_e,\varphi_j,\varphi_e) \hbox{ real},
\end{eqnarray}
and introduces the notation
\begin{eqnarray}
& & {\hskip -12.0 truemm}
\Phi_{12}=\varphi_1-\varphi_2-\varphi_e + \frac{\pi}{2},\
\Phi_{43}=\varphi_4-\varphi_3-\varphi_e + \frac{\pi}{2},\
\end{eqnarray}
the system (\ref{eqFWM}) becomes
\begin{eqnarray}
& & {\hskip -13.0 truemm}
\left\lbrace
\begin{array}{ll}
\displaystyle{
\partial_z       M_1 = +       M_2 M_e       \cos \Phi_{12},\ 
\partial_z \varphi_1 = - \frac{M_2 M_e}{M_1} \sin \Phi_{12},\ 
}\\ \displaystyle{
\partial_z       M_2 = -       M_1 M_e       \cos \Phi_{12},\ 
\partial_z \varphi_2 = - \frac{M_1 M_e}{M_2} \sin \Phi_{12},\ 
}\\ \displaystyle{
\partial_z       M_4 = -       M_3 M_e       \cos \Phi_{43},\ 
\partial_z \varphi_4 = + \frac{M_3 M_e}{M_4} \sin \Phi_{43},\ 
}\\ \displaystyle{
\partial_z       M_3 = +       M_4 M_e       \cos \Phi_{43},\ 
\partial_z \varphi_3 = + \frac{M_4 M_e}{M_3} \sin \Phi_{43}.
}
\end{array}
\right. 
\label{eqFWM_modulus_arg}
\end{eqnarray}
It is then convenient to introduce the first integrals
\begin{eqnarray}
& & 
f_{12}^2(t)=\mod{A_1}^2+\mod{A_2}^2,\
f_{43}^2(t)=\mod{A_4}^2+\mod{A_3}^2,\
\end{eqnarray}
and to compute the $z$-evolution of the two functions
\begin{eqnarray}
& & 
v_{12}=\mod{A_1}^2-\mod{A_2}^2,\
v_{43}=\mod{A_4}^2-\mod{A_3}^2.
\end{eqnarray}
One finds
\begin{eqnarray}
& & 
\partial_z v_{12}= + 4 M_1 M_2 M_e \cos \Phi_{12},\
 \nonumber\\ & &
\partial_z v_{43}= - 4 M_4 M_3 M_e \cos \Phi_{43},\ 
\end{eqnarray}
and, by elimination of $M_j$,
\begin{eqnarray}
& & 
\left(\partial_z v_{12}\right)^2=
4 \left(f_{12}^4 - v_{12}^2\right) \mod{\ee}^2 \cos^2 \Phi_{12},\
\nonumber\\ & &
\left(\partial_z v_{43}\right)^2=
4 \left(f_{43}^4 - v_{43}^2\right) \mod{\ee}^2 \cos^2 \Phi_{43}.
\label{eqvijode1}
\end{eqnarray}
If one defines two functions $u_{12}(z,t),u_{43}(z,t)$
by the relations
\begin{eqnarray}
& & {\hskip -12.0 truemm}
\mod{\ee}^2 \cos^2 \Phi_{12} =\left(\partial_z u_{12}\right)^2,\
\mod{\ee}^2 \cos^2 \Phi_{43} =\left(\partial_z u_{43}\right)^2,\
\end{eqnarray}
the two equations (\ref{eqvijode1})
can be integrated explicitly in terms of the two variables $u_{12},u_{43}$,
\begin{eqnarray}
& & 
v_{12}=  - f_{12}^2 \cos (2 (u_{12} - c_{12})),\
\nonumber\\ & &
v_{43}=  - f_{43}^2 \cos (2 (u_{43} + c_{43})),\
\label{eqvijsol}
\end{eqnarray}
with $c_{12}$ and $c_{43}$ arbitrary functions of $t$.
Basic trigonometry then yields
\begin{eqnarray}
& & {\hskip -13.0 truemm}
\left\lbrace
\begin{array}{ll}
\displaystyle{
A_1=+f_{12} \sin (u_{12} -c_{12}) e^{i \varphi_1},\
}\\ \displaystyle{
A_2=+f_{12} \cos (u_{12} -c_{12}) e^{i \varphi_2},\
}\\ \displaystyle{
A_4=-f_{43} \sin (u_{43} +c_{43}) e^{i \varphi_4},\
}\\ \displaystyle{
A_3=+f_{43} \cos (u_{43} +c_{43}) e^{i \varphi_3}.
}
\end{array}
\right. 
\label{eqFWMAj_cjk}
\end{eqnarray}

We have not succeeded to similarly integrate the equations
for $\varphi_j$ in (\ref{eqFWM_modulus_arg})
without any additional assumption.
Let us therefore assume,
as was done in \cite{JBH_JOSAB,BKK,BKMPR},
that these four equations for the
spatial evolution of $\varphi_j$ identically vanish,
i.e.~that 
$\sin \Phi_{12}=\sin \Phi_{43}=0$ 
and the phases $\varphi_j$ are independent of $z$, 
\begin{eqnarray}
& & 
\partial_z \varphi_j=0,\ j=1,2,3,4,\
\nonumber\\ & &
\Phi_{12}=n_{12} \pi,\
\Phi_{43}=n_{43} \pi,\
n_{12} \hbox{ and }
n_{43} \hbox{ integers},\
\nonumber\\ & &
\partial_z \varphi_e=0,\ 
\end{eqnarray}
and for convenience let us redefine the solution (\ref{eqFWMAj_cjk})
as
\begin{eqnarray}
& & {\hskip -11.0 truemm}
\left\lbrace
\begin{array}{ll}
\displaystyle{
\ee=\left(\partial_z u\right) e^{\displaystyle{i \varphi_e}},\ 
}\\ \displaystyle{
A_1=+f_{12} \sin (s_{12} (u -C_{12})) e^{\displaystyle{i\varphi_1}},\ 
}\\ \displaystyle{
A_2=+f_{12} \cos (s_{12} (u -C_{12})) e^{\displaystyle{i\varphi_2}},\ 
}\\ \displaystyle{ 
A_4=-f_{43} \sin (s_{43} (u +C_{43})) e^{\displaystyle{i\varphi_4}},\ 
}\\ \displaystyle{
A_3=+f_{43} \cos (s_{43} (u +C_{43})) e^{\displaystyle{i\varphi_3}},\ 
}\\ \displaystyle{ 
\Phi_{12} \equiv \varphi_1-\varphi_2-\varphi_e + \frac{\pi}{2} = n_{12} \pi,\ 
s_{12}=(-1)^{n_{12}},\
}\\ \displaystyle{
\Phi_{43} \equiv \varphi_4-\varphi_3-\varphi_e + \frac{\pi}{2} = n_{43} \pi,\
s_{43}=(-1)^{n_{43}},\ 
}\\ \displaystyle{
\intIm=\frac{1}{2} 
\left( f_{12}^2 \sin 2 (u -C_{12}) 
      -f_{43}^2 \sin 2 (u +C_{43}) \right)
}\\ \displaystyle{\phantom{123456}
        \times e^{\displaystyle{i(\varphi_e - \pi/2)}},
}\\ \displaystyle{
\intId=f_{12}^2 \cos 2 (u -C_{12}) 
      +f_{43}^2 \cos 2 (u +C_{43}),
}\\ \displaystyle{
n_{12}, n_{43} \in \mathcal{Z}.
}
\end{array}
\right.
\label{eqFWMparamuSG} 
\end{eqnarray}

The last complex equation to be enforced (\ref{eqgrating_t})
is equivalent to the two real equations
\begin{eqnarray}
& & {\hskip -11.0 truemm}
\partial_z \partial_t u 
+ \frac{1}{\tau} \partial_z u - K \sin(2 u + \alpha)=0,\
\nonumber \\ & & {\hskip -11.0 truemm}
K e^{i \alpha} = \frac{\gnl \sin g}{2}
\left(f_{12}^2 e^{- 2 i C_{12}} - f_{43}^2 e^{ 2 i C_{43}}\right),\
\label{eqdampedSG0} 
\\
& & {\hskip -11.0 truemm}
(\partial_z u) (\partial_t \varphi_e) + (\cotg g) K \sin(2 u + \alpha)=0,\
\gamma=\mod{\gamma} e^{i g}.
\label{eqdampedSGsecond} 
\end{eqnarray}

If $\partial_t \varphi_e\not=0$, 
the ODE (\ref{eqdampedSGsecond}) (with $t$ as a parameter)
integrates as
\begin{eqnarray}
& & {\hskip -8.0 truemm}
\left\lbrace
\begin{array}{ll}
\displaystyle{
\cos(2u+\alpha)
 =+\tanh 2\left(\frac{K(t)\cotg g}{\partial_t \varphi_e} (z-z_0(t)) \right),\
}\\ \displaystyle{
\sin(2u+\alpha)
 =-\sech 2\left(\frac{K(t)\cotg g}{\partial_t \varphi_e} (z-z_0(t)) \right),\
}
\end{array}
\right. 
\end{eqnarray}
then the equation (\ref{eqdampedSG0}) restricts this solution to
\begin{eqnarray}
& & {\hskip -8.0 truemm}
\varphi_e=-\frac{\cotg g}{\tau} (t-t_0),\
\partial_z u=-K \tau \sech 2 K \tau (z-z_0),\
\\ & & {\hskip -8.0 truemm}
\cos(2u+\alpha)=-\tanh 2 K \tau (z-z_0),\
\nonumber \\ & & {\hskip -8.0 truemm}
\sin(2u+\alpha)=-\sech 2 K \tau (z-z_0),\
\end{eqnarray}
in which $K,t_0,z_0$ are arbitrary constants.
This solution can also be viewed 
as the general solution of the
reduction $\intIm / \ee=$ complex constant of the intrinsic system
(\ref{eqFWM555}),
\begin{eqnarray}
& & {\hskip -6.0 truemm}
\forall \tau,\gamma:\
\left\lbrace
\begin{array}{ll}
\displaystyle{
\intId= - \frac{2 K}{\mod{\gamma} \sin g} \tanh 2 K \tau (z-z_0),
}\\ \displaystyle{
\ee=-i \mod{\gamma} (\sin g) \tau \intIm
}\\ \displaystyle{\phantom{12}
=-e^{-i (\cotg g)(t-t_0)/\tau} K \tau \sech 2 K \tau (z-z_0),
}
\label{eqeeIm}
\end{array}
\right. 
\end{eqnarray}
in which the wave number $K$ is arbitrary.
Very similar to \cite[Eq.~(23)]{CB2009a},
this solution is however new and it depends on both space and time.

If $\partial_t \varphi_e=0$, then $\gamma$ must be purely imaginary
\begin{eqnarray}
& &
\partial_t \varphi_e=0,\
\cos g=0,\ 
\end{eqnarray}
this defines the already investigated damped sine-Gordon equation.

The result of the above computation can be summarized as follows.
Under the three assumptions that
the phases of each $A_j$ are independent of $z$,
the phase of $\ee$ is constant,
and $\gamma$ is purely imaginary,
one obtains a solution 
of the system
(\ref{eqFWM})--(\ref{eqdefIm})
represented as (\ref{eqFWMparamuSG}),
in terms of the real solution $u$ of a damped sine-Gordon equation
(\ref{eqdampedSG0}) (with $\sin g=1$). 
The representation (\ref{eqFWMparamuSG}) displays the invariance
$(1,2,3,4,\partial_z,u) \to (4,3,2,1,-\partial_z,-u)$
and depends on six arbitrary real functions of $t$ 
($f_{12}$, $f_{43}$, $C_{12}$, $C_{43}$,
$\varphi_1+\varphi_2$, $\varphi_4+\varphi_3$) 
and one arbitrary real constant (the phase $\varphi_e$).

\section{$\Re(\gamma)=0$. From real damped sine-Gordon to CGL3}
\label{section_multiscale}

It is a classical result \cite{TaniutiYajima1969}
that the nonlinear Schr\"odinger equation (NLS)
can be derived from the sine-Gordon equation by 
a reductive perturbation method,
see details in e.g.~\cite{DauxoisPeyrard}.
When applied to the real damped sine-Gordon equation 
(\ref{eqdampedSG0}),
this method yields a 
complex cubic Ginzburg-Landau 
equation which we now derive.
Consider the damped sine-Gordon equation  (\ref{eqdampedSG0})
\begin{eqnarray}
& & {\hskip - 5.0 truemm}
E\equiv
\partial_t \partial_z u + \frac{1}{\tau} \partial_z u 
 - K(t) \sin (2 u + \alpha(t))=0,
\end{eqnarray}
in which $u(z,t)$, $K(t)$, $\alpha(t)$ and $\tau$ are real.
 
Following the classical derivation of NLS from the sine-Gordon equation
\cite{TaniutiYajima1969,DauxoisPeyrard}, 
we define a multiscale expansion in which $u$ is of order
$\varepsilon$, while $K(t)$ is of order one,
\begin{eqnarray}
& & {\hskip - 5.0 truemm}
\left\lbrace
\begin{array}{ll}
\displaystyle{
u(z,t)+\frac{\alpha(t)}{2}
}\\ \displaystyle{\phantom{123}
=\varepsilon \sum_{j=0}^{+\infty}
\varepsilon^j 
\varphi_j(z,\varepsilon z,\dots,\varepsilon^k z,\dots,
          t,\dots,\varepsilon^k t,\dots),
}\\ \displaystyle{
K(t)=\sum_{j=0}^{+\infty} \varepsilon^j 
      K_j(\varepsilon^j t,\dots,\varepsilon^k t,\dots),\
\tau=\hbox{ unchanged},
}\\ \displaystyle{
E=\varepsilon \sum_{j=0}^{+\infty}
\varepsilon^j E_j,
}
\end{array}
\right.
\label{eqmultiDef}
\end{eqnarray}
and, after renaming the scaled independent variables as
\begin{eqnarray}
& &
\varepsilon^k z=Z_k,\ \varepsilon^k t=T_k,\ 
\label{eqmultiZkTk}
\end{eqnarray}
one requires each coefficient $E_j$ to vanish.

The zero-th order
\begin{eqnarray}
& & {\hskip - 5.0 truemm}
L \varphi_0=0,\
L \equiv
\partial_{T_0} \partial_{Z_0} + \frac{1}{\tau}\partial_{Z_0}
-2 K_0(T_0,\dots),
\label{eqmultiOrder0}
\end{eqnarray}
admits the plane wave-type complex solution
\begin{eqnarray}
& &
\varphi_0=A(Z_1,Z_2,T_1,T_2,\dots)
           e^{i(q Z_0-F(T_0,\dots))},\
\label{eqmultiOrder0OneWave}
\end{eqnarray} 
in which the complex constant $q$
and the complex function $F$ obey the dispersion relation
\begin{eqnarray}
& &
i \frac{q}{\tau} + q \frac{\partial F} {\partial {T_0}}
- 2 K_0=0.
\label{eqdSGtoNLS_dispersionc}
\end{eqnarray}
Since $K_0$ may depend on $T_0$,
it is convenient to introduce the primitive $Q_0$ of $K_0$
and to represent the dispersion relation by its integrated form
\begin{eqnarray}
& &
F=-i \frac{T_0}{\tau} + 2 \frac{Q_0}{q},\
\frac{\partial Q_0} {\partial T_0}=K_0.
\label{eqdSGtoNLS_dispersionc_integr}
\end{eqnarray}
The physical solution of
(\ref{eqmultiOrder0}) is then chosen as the real part
of the above complex plane wave
\begin{eqnarray}
& &
\varphi_0=A(Z_1,Z_2,T_1,T_2,\dots) e^{\Phi_0} + \hbox{c.c.},\
\nonumber\\ & &
\Phi_0= i q Z_0 - \frac{T_0}{\tau} - 2 i \frac{Q_0}{q}.
\end{eqnarray}

The first order equation, which defines the evolution of $\varphi_1$,
\begin{eqnarray}
& & {\hskip - 5.0 truemm}
L \varphi_1=-       G_{1}  e^{          \Phi_0}
        - \overline{G_{1}} e^{\overline{\Phi}_0},\
\nonumber\\ & & {\hskip - 5.0 truemm}
G_{1} \equiv
- \frac{2 i K_0}{q} \frac{\partial A}{\partial Z_1}
+ i q               \frac{\partial A}{\partial T_1}
-2 \left(K_1 - \frac{\partial Q_0}{\partial T_1}\right) A,
\end{eqnarray}
requires the vanishing of $G_1$ in order to avoid $\varphi_1$ to diverge.
This defines two complex conjugate linear PDEs for 
$A(Z_1,T_1)$ and $\overline{A}$,
and the solution of this first order is
\begin{eqnarray}
& &
\left\lbrace
\begin{array}{ll}
\displaystyle{
\varphi_1=0,
}\\ \displaystyle{
A=a(Z_1-\vg T_1,Z_2-\vg T_2,T_2,\dots) e^{\Phi_1},\
}\\ \displaystyle{
\Phi_1=-2 i \frac{Q_1-Q_0}{q},\
\vg=-\frac{2 K_0}{q^2},\
\frac{\partial Q_1} {\partial T_1}=K_1,
}
\end{array}
\right.
\end{eqnarray}
in which the complex function of integration $a$ is 
to be determined.

Since the group velocity $\vg$ is generically complex,
let us introduce the two complex conjugate independent variables $X_1,Y_1$,
\begin{eqnarray}
& &
X_1=Z_1-\vg T_1,\
Y_1=\overline{X_1}=Z_1-\vc T_1.
\end{eqnarray}
The second order equation similarly defines the evolution of $\varphi_2$,
\begin{eqnarray}
& & {\hskip - 7.0 truemm}
L \varphi_2=
-\frac{4}{3} K_0 
\left(e^{3(          \Phi_0 +\Phi_1)} a^3
     +e^{3(\overline{\Phi_0}+\overline{\Phi_1)}} 
     \bar{a}^3
\right)
\nonumber\\ & &  {\hskip - 7.0 truemm} \phantom{1234}
 -           q G_{2}  e^{          \Phi_0 +\Phi_1}
 - \overline{q G_{2}} e^{\overline{\Phi_0}+\overline{\Phi_1}},\
\\ 
& & {\hskip - 7.0 truemm}
G_{2} \equiv
i \frac{\partial a}{\partial T_2}
- 2 \frac{i K_1}{q^2} \frac{\partial a}{\partial X_1}
+\frac{2 K_0}{q^3} \frac{\partial^2 a}{\partial^2 X_1} 
\nonumber\\ & &  {\hskip - 7.0 truemm} \phantom{1234}
+4 \frac{K_0}{q} e^{2 \Re(\Phi_0+\Phi_1)} \mod{a}^2 a
- \frac{2}{q} \left(K_2-\frac{\partial Q_1}{\partial T_2}\right) a,
\label{eqCGL3brut}
\end{eqnarray}
and the cancellation of the secular terms requires $G_2$ to vanish,
which defines two complex conjugate nonlinear PDEs for 
$a(X_1,T_2)$ and $\bar a(Y_1,T_2)$
and yields the value
\begin{eqnarray}
& &
\varphi_2=\frac{2 \tau K_0}{24 \tau K_0 - 9 i q} a e^{3(\Phi_0+\Phi_1)}
+ \hbox{ c.c.}
\end{eqnarray}

Therefore, under the reductive perturbation method,
the damped sine-Gordon equation (\ref{eqdampedSG0})
generically yields the complex PDE $G_2=0$ Eq.~(\ref{eqCGL3brut}),
in which $K_0,K_1,K_2$ depend on $T_2$.

In the pure sine-Gordon limit $1/\tau=0, K(t)=k_0=$ constant,
with $q$ real,
one checks that the PDE $G_2=0$ reduces to the nonlinear Schr\"odinger equation,
\begin{eqnarray}
& &
\frac{1}{\tau}=0,\ K(t)=k_0,\ q \hbox{ real}:\
\nonumber\\ & &
i \frac{\partial a}{\partial T_2}
+\frac{2 k_0}{q^3} \frac{\partial^2 a}{\partial^2 X_1} 
+4 \frac{k_0}{q} \mod{a}^2 a=0.
\label{eqNLSbrut}
\end{eqnarray}

In the generic case ($q$ complex),
the PDE $G_2=0$ Eq.~(\ref{eqCGL3brut}) would be identical to
the cubic complex Ginzburg-Landau equation (CGL3)
if its coefficients were independent of $T_2$.
Let us therefore try to get rid of this dependence on $T_2$
by performing the transformation
\begin{eqnarray}
& &
a(X_1,T_2)=\psi(\xi,\eta) e^{\lambda(T_2)},\ 
\nonumber\\ & &
\xi=X_1- f_1(T_2),\ \eta=f_2(T_2),
\label{eqCGLtransfo}
\end{eqnarray}
in which the complex functions $f_1,\lambda$ 
and the real function $f_2$ can be freely chosen.
The best one can achieve is to concentrate the dependence on $T_2$
in only one coefficient, e.g.~the gain or loss term.
Then the functions of the transformation are the following,
\begin{eqnarray}
& &
\frac{\D f_1}{\D T_2}=-2 \frac{K_1}{q^2},\
\frac{\D f_2}{\D T_2}=K_0,\
\nonumber\\ & &
\lambda=\frac{2 i}{q} \int K_1 \D T_1
        -2 i \frac{\Re(q)}{\mod{q}^2} \int K_2 \D T_2.
\end{eqnarray} 

The final CGLE is
\begin{eqnarray}
& &
i \frac{\partial \psi}{\partial \eta}
+\frac{2}{q^3} \frac{\partial^2 \psi}{\partial^2 \xi} 
+\frac{4}{q} e^{\displaystyle{-2 \frac{T_0}{\tau} - 2 \Im(q) Z_0}} \mod{\psi}^2 \psi
\nonumber\\ & & \phantom{1234}
+ 2 i \frac{\Im(q)}{\mod{q}^2} \frac{K_2}{K_0} \psi=0.
\label{eqCGL3-with-t2}
\end{eqnarray} 
in which the coefficient $K_2/K_0$ depends on $T_2$ and the other
coefficients are complex constants.
Under the condition that $K_2/K_0$ be independent of $T_2$,
the above PDE (\ref{eqCGL3-with-t2}) is then identical to the CGL3 equation.

We want to emphasize here that the CGLE (\ref{eqCGL3-with-t2}) includes only 
the longitudinal space coordinate $\xi$ for the variable $\psi$. 
It does not contain any transverse spatial coordinates.

We thus obtain the CGLE for describing 
the dynamics of the FWM in a nonlocal medium with a dissipative term, 
where the dependent 
variable is the envelope of the potential $u$.
With the definition 
$\mod{\ee}=\partial_z u$ and the multiscale expansion 
$\mod{\ee}=\varepsilon \sum_{j=0}^{+\infty} \varepsilon^j \ee_j$,
one can obtain the expression which connects the value $u$ 
with the envelope of the spatial distribution of the nonlinearity,
\begin{eqnarray}
& & {\hskip - 5.0 truemm}
\ee_0
=\frac{\partial \varepsilon_0}{\partial Z_0} 
\nonumber\\ & &  {\hskip - 5.0 truemm} \phantom{123}
= i q \psi(\xi,\eta)
 e^{i(q Z_0- \frac{2}{q} Q_0-2 \frac{\Re(q)}{|q|^2} Q_2+ \frac{i T_0}{\tau})}
+ \hbox{ c.c.}
\label{eqeeuSG}
\end{eqnarray} 

Taking into account the equation (\ref{eqeeIm})
which connects $\ee$ and the intensity field $\intIm$,
we obtain that both the nonlinearity spatial shape and the time behavior 
of the intensity pattern are the same spatiotemporal 
distribution, 
where the magnitudes $\intIm$ and $\ee$ only differ by a constant. 
The complex sign "$i$" means that the functions $\intIm$ and $\ee$ 
have a relative shift 
in the spatial coordinate. 

Thus the CGLE (\ref{eqCGL3-with-t2}) together with the Eqs.~(\ref{eqeeuSG}) 
and (\ref{eqeeIm}) describe the statiotemporal dynamics for 
the physical field $\intIm$ (the interference pattern), 
the parameter of the phase transition $\ee$ and the potential $u$.

\section{Conclusion}
\label{section_conclusion}

We have obtained the complex Ginzburg-Landau equation from the nonlinear systems of the dynamical four-wave mixing that includes degenerate wave-coupling in a cubic nonlinear medium which has both nonlocal and relaxation response.
The obtained CGLE is just the cubic one when the response is purely nonlocal, i.e. there is the energy transfer only between the interacting waves but no phase transfer. 
In this case the initial FWM system is reduced to a damped sine-Gordon equation containing the first derivative on the spatial longitudinal coordinate $z$. 
We show that by applying the reductive perturbation method, the real damped sine-Gordon equation reduces to the cubic CGLE, except for a loss/gain coefficient dependent on time. 
The cubic CGLE describes the dynamics of the formation of localized states (intensity patterns) along longitude z-direction in bulk nonlinear medium.

The interest is to apply the reductive perturbation method to the generic system with the complex response. 
We show the initial generic complex FWM is reduced to the intrinsic system, which has three dependent variables ($I_m$, $\ee$ are complex ones, and $I_d$ is real).
The intrinsic system has a form very similar to the complex Maxwell-Bloch system. 
It coincides completely with the Maxwell-Bloch system when at the same time the response is purely nonlocal the time relaxation is absent. 
In optics there exists an example of reduction of the Maxwell-Bloch system to the CGLE, which describes the formation of transverse mode structures in lasers \cite{Staliunas}, but they are derived in a high-order approximation.

Till nowadays a number of solutions of the CGLE have been found including stable localized patterns \cite{vS2003}, i.e.~dissipative solitons.
These solutions may find applications in the dissipative FWM system.
They have great potential for practical use in photonics by applying wave-coupling with a nonlocal medium.

\begin{acknowledgments}
This is a real pleasure to warmly acknowledge the financial support of the
Max-Planck-Institut f\"ur Physik komplexer Systeme.

\end{acknowledgments}


\vfill\eject

\end{document}